\documentclass[prl,showpacs,twocolumn]{revtex4}%
\usepackage{amsfonts}
\usepackage{amsmath}
\usepackage{amssymb}
\usepackage{graphicx}%
\begin{document}
\title{Low compressible noble metal carbides with rock-salt structure: \textit{ab
initio} total energy calculations of the elastic stability }
\author{Chang-Zeng Fan, Song-Yan Zeng}
\affiliation{Department of Material Science and Engineering, Harbin
Institute of Technology, Harbin 150001, China}
\author{Zai-Ji Zhan, Ri-Ping Liu, Wen-Kui Wang}
\affiliation{Key Laboratory of Metastable Material Science and Technology, Yanshan
University, Qinhuangdao 066004, China}
\author{Ping Zhang }
\affiliation{Institute of Applied Physics and Computational Mathematics, Beijing 100088, China}
\author{Yu-Gui Yao}
\affiliation{Institute of Physics, Chinese Academy of Sciences, Beijing 100080, China}

\begin{abstract}
We have systematically studied the mechanical stability of all noble metal
carbides with the rock-salt structure by calculating their elastic constants
within the density function theory scheme. It was found that only four
carbides (RuC, PdC, AgC and PtC) are mechanically stable. In particular, we
have shown that RuC, PdC, and PtC have very high bulk modulus, which has been
remarkably observed by the most recent experiment for the case of PtC. From
the calculated density of states, we can conclude that these compounds are
metallic, like the conventional group IV and group V transition metal carbides.

\end{abstract}
\pacs{81.05.Zx, 62.20.Dc, 71.20.Be, 61.66.Fn}
\maketitle

Transition metal carbides (TMCs) have attracted much attention due to their
excellent physical, chemical and mechanical properties.\cite{stor,toth} In the
past, most work was focused on the group IV and group V
TMCs,\cite{blah,schw,jhi,sahn,chen} while noble metal carbides have been only
studied spectroscopically by few researchers\cite{lang,melo,lind} because the
crystalline samples are hard to obtain. Very recently, shortly after the
synthesis of a novel noble-metal nitride, PtN,\cite{greg} Shigeaki Ono
\textit{et al}.\cite{ono} have succeeded for the first time in synthesizing
platinum carbide (PtC) at high-pressure and high-temperature by the
laser-heated diamond anvil cell technique. According to their report, the new
PtC has a rock-salt (RS) structure, with a high bulk modulus of 301 ($\pm1$5) GPa.

While the structual and electronic properties of earlier-synthesized PtN have
been extensively studied in recent
theoretical\cite{sahu,uddin,yu1,fan1,yu2,kanoun,yu3,patil,fan2,young1,young2}
and experimental\cite{crow} work, very few attention has been payed to the PtC
system. Recently, Li \textit{et al}.\cite{lily} have given a first-principles
calculation of PtC. They preferred PtC crytallized in zinc-blende (ZB)
structure since their results showed that the RS-PtC is mechanically unstable.
In contrast, after a systematic calculation, and based on two separate DFT
schems, we found that the RS-PtC is mechanically stable\cite{fan3} and the
compressibility behavior of RS-PtC is more comparable with that of the
experiment. To illustrate, here we show in Fig. 1%
\begin{figure}[tbp]
\begin{center}
\includegraphics[width=0.90\linewidth]{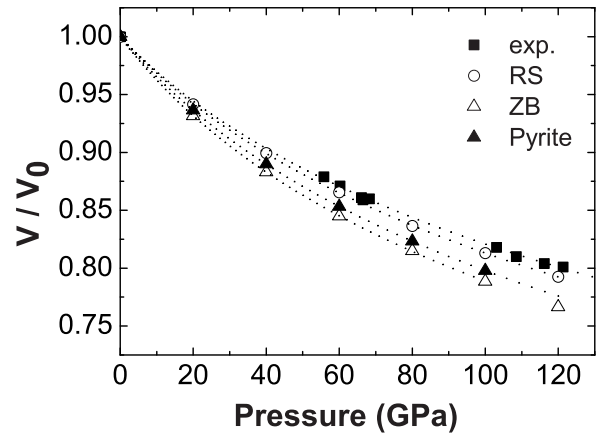}
\end{center}
\caption{Calculated Pressure-Volume data (average LDA and GGA results) of PtC
with RS and ZB structure, PtC$_{2}$ with Pyrite structure and experimental
data. Dashed curves are the third-order Birch-Murnaghan equation fit with
$B$=289 (239, 257, 301) GPa and $B^{\prime}$=5.0 (4.7, 4.6 5.3) for the RS
(ZB, Pyrite, exp.).
}
\label{fig1}
\end{figure}
the calculated pressure-volume relation for both ZB- and RS-PtC. For
comparison, the experimental data\cite{ono} are also plotted. Clearly, one can
see that the calculated compressibility ratio for RS-structure is in good
agreement with the experimental result while it is not so for ZB-structure.
Therefore, our results firmly show that the observed phase of PtC under
high-pressure is rocksalt, although it is difficult to distinguish the
structures of RS from ZB only by X-ray diffraction measurement\cite{Ono2}. We
have also found\cite{fan3} that the total energy of the ZB structure is lower
than that of the RS structure (about 1 eV/unit cell). From this sense, the
synthesized PtC in Ref.[12] is metastable. In addition, we have further
calculated the elastic constants of PtC$_{2}$ with the fluorite and pyrite
structures, since the former structure was once proposed to explain the high
bulk modulus of platinum nitride and the latter structure has been recently
observed\cite{crow} in platinum nitride. Our calculation revealed that the
fluorite PtC$_{2}$ is mechanically unstable. On the other hand, the pyrite
PtC$_{2}$ is calculated to be mechanically stable but with a low bulk modulus
(B=236 GPa, B'=4.7), much lower than that of the experiment.\cite{ono} What's
more, there also exsits a big discrepancy between the calculated
pressure-volume data of pyrite PtC$_{2}$ and those of experiments (see Fig.1).
Therefore, we excluded the possibility of existence of zinc-blende or pyrite
structure in the experiment.\cite{ono}

To appreciate the consistency of our calculation with the experimental
observation on PtC and its very high bulk modulus, in this paper we have gone
beyond to search in a systematic way for other possible mechanically stable
noble-metal carbides with the RS structure by calculating the relevant elastic
constants. Surprisingly, we have found that RuC has even higher bulk modulus
of $\sim$370 GPa. Among the whole noble metal (Ru, Rh, Pd, Ag, Os, Ir, Pt, and
Au), Ag and Pd were also found to form mechanically stable carbides. Together
with RuC and PtC, they form a new family of low compressible compounds. The
partial density of states (PDOS) of RuC, PdC, AgC, and PtC were calculated,
which revealed that the hybridization of metal 5\textit{d}-like and C
2\textit{p}-like states is responsible for the strong covalent bonding of
these compounds. The PDOS also revealed that these compounds are of metallic character.

The first-principles calculations were performed by the plane-wave
pseudo-potential (PW-PP) method as implemented by CASTEP code\cite{sega}. The
ultrasoft pseudopotential (USPP)\cite{vand} was used to describe the
interaction between ions and electrons. The exchange and correlation terms
were described by the local-density approximation (LDA)\cite{cepe} and
generalized gradient approximation (GGA)\cite{perd}. For the Brillouin-zone
sampling, the Monkhorst-Pack (MP) scheme\cite{monk} with a mesh of (12, 12,12)
was adopted. We chose Ecut =550 eV for all the calculations. It was tested
that with even more strict parameters the total energy can be converged within
0.003 eV/atom for all the cared systems. For the mechanical stable phases, the
elastic constants were recalculated with the full-potential linearized
augmented plane-wave (LAPW) method\cite{blah2} (implemented by WIEN2K),
wherein a non-overlapping muffin-tin (MT) sphere radius of 2.2 (2.1, 2.3, 2.3)
a.u. for Ru (Pd, Ag, Pt) and 1.7 (1.5, 1.7, 1.7) a.u. for C atoms in RuC (PdC,
AgC, PtC) compound was employed. The valence wave functions inside the MT
spheres were expanded into spherical harmonics up to \textit{l}=10 and the
potential up to \textit{l}=4 for the 4\textit{d} carbides and \textit{l}=6 for
the 5\textit{d} carbide (PtC). The energy that separates the core and valence
states was set to be -6.0 \textit{R}$_{y}$ for the 4\textit{d} carbides and
-7.0 \textit{R}$_{y}$ for the 5\textit{d} carbides. Here, an APW+lo\cite{sjos}
type basis set for certain electronic states was also used. We used 5000 k
points for sampling the Brillouin zone and \textit{R}$_{mt}\times$%
\textit{K}$_{max}$ was taken as 8.0 (where \textit{R}$_{mt}$ is the muffin-tin
radius and \textit{K}$_{max}$ is the plane-wave cutoff). A fully relativistic
calculation was performed for core states, whereas the valence states were
treated in a scalar relativistic scheme.

After getting the equilibrium geometry configures, in the PW-PP method we
applied the so-called \textquotedblleft stress-strain\textquotedblright%
\ method to obtain the elastic constants. For the cubic crystal, there are
only three different symmetry elements (\textit{c}$_{11}$, \textit{c}$_{12}$
and \textit{c}$_{44}$). A single strain with non-zero first and fourth
components (\textit{xx} and \textit{yz}) can give stresses relating to all
three of these coefficients, yielding a very efficient method for obtaining
elastic constants for the cubic system. The elastic constants can be obtained
from the stress-strain relations and then the bulk modulus is obtained from
the elastic constants by equation \textit{B}=(\textit{c}$_{11}$+2\textit{c}%
$_{12}$)/3. In the LAPW method another scheme was adopted (For details read
document in WIEN2K package). In this scheme, three types of deformation were
performed to produce three linear equations for \textit{c}$_{11}$,
\textit{c}$_{12}$, and \textit{c}$_{44}$.

\begin{table*}
\caption{The calculated equilibrium lattice parameters $a$(\AA ),
elastic constants $c_{ij}$ (GPa), bulk modulus $B$ (GPa), tetragonal
shear modulus
$G^{\prime\prime}$ (GPa), polycrystalline shear modulus $G$=($c_{11}$-$c_{12}%
$+3$c_{44}$)/5 (GPa), Young's modulus $E$ (GPa) and Poisson's ratio
of four
mechanically stable noble metal carbides. }%
\begin{ruledtabular}
\begin{tabular}
[c]{ccccccccccc}
&  & $a$ & $c_{11}$ & $c_{44}$ & $c_{12}$ & $G^{\prime\prime}$ & $B$
& $G$ & $E$ & $v$\\\hline RuC & PW-LDA & 4.236 & 604.9 & 0.9 & 252.8
& 176.1 & 370.2 & 71.0 & 200.2 &
0.41\\
& PW-GGA & 4.299 & 504.5 & 8.8 & 233.8 & 135.4 & 324.0 & 59.4 & 168.0 & 0.41\\
& Ave. & 4.268 & 554.7 & 4.9 & 243.3 & 155.8 & 347.1 & 65.2 & 184.1 & 0.41\\
& LAPW & 4.327 & 487.7 & 21.6 & 226.7 & 130.5 & 313.7 & 65.2 & 182.9 & 0.40\\
PdC & PW-LDA & 4.340 & 411.5 & 48.4 & 214.6 & 98.5 & 280.2 & 68.4 &
166.4 &
0.22\\
& PW-GGA & 4.430 & 297.9 & 43.8 & 176.2 & 60.9 & 216.8 & 50.6 & 140.8 & 0.39\\
& Ave. & 4.385 & 354.7 & 46.1 & 195.4 & 79.7 & 248.5 & 59.5 & 103.6 & 0.31\\
& LAPW & 4.433 & 278.3 & 41.8 & 189.6 & 44.4 & 219.2 & 42.8 & 120.5 & 0.41\\
AgC & PW-LDA & 4.543 & 246.0 & 17.6 & 147.7 & 49.2 & 180.4 & 30.2 &
85.8 &
0.42\\
& PW-GGA & 4.665 & 185.3 & 14.3 & 113.0 & 36.2 & 137.0 & 23.1 & 65.6 & 0.42\\
& Ave. & 4.604 & 215.7 & 16.0 & 130.4 & 42.7 & 158.8 & 26.7 & 75.7 & 0.42\\
& LAPW & 4.657 & 189.8 & 8.2 & 113.2 & 38.3 & 138.7 & 26.7 & 75.3 & 0.41\\
PtC & PW-LDA & 4.425 & 373.6 & 47.0 & 284.0 & 44.8 & 313.9 & 46.1 &
131.8 &
0.43\\
& PW-GGA & 4.506 & 277.8 & 51.7 & 256.4 & 10.7 & 263.5 & 35.3 & 101.4 & 0.44\\
& Ave. & 4.466 & 325.7 & 49.4 & 270.2 & 27.8 & 288.7 & 40.7 & 116.6 & 0.44\\
& LAPW & 4.489 & 272.8 & 11.9 & 255.5 & 8.7 & 261.3 & 10.6 & 31.4 & 0.48\\
& Exp.(Ref.[12]) & 4.814 &  &  &  &  & 301$\pm1$5 &  &  &
\\
\end{tabular}
\end{ruledtabular}
\end{table*}

Table I lists the calculated elastic properties of the mechanically
stable noble metal carbides. One can see that the results from the
PW-PP method and those from the LAPW method fit very well. An
exception is the amplitude of \textit{c}$_{44}$\textit{ }in PtC and
RuC; the inconsistency between the two methods remains unclear.
However, since the amplitude of \textit{c}$_{44}$ does not affect
the value of the bulk modulus of these compounds, we do not care
about it at present. As it is well known that in first-principles
calculations the LDA (GGA) usually underestimates (overestimates)
the lattice constant and overestimates (underestimates) the bulk
modulus, hence we adopt in the PW-PP method to use the average of
LDA and GGA values as the re-scaled estimates. The responses of the
crystal under hydrostatic pressure, rhombohedral, and tetragonal
distortions are measured by the bulk modulus $B$, the normal shear
modulus $G^{\prime}=c_{44}$, and the tetragonal shear modulus
$G^{\prime\prime}=$($c_{11}$-$c_{12}$)/2, respectively. The
mechanical stability of a crystal requires the strain energy to be
positive, which for cubic crystal implies\cite{Nye}
\begin{equation}
c_{44}>0,c_{11}>|c_{12}|,c_{11}+2c_{12}>0. \tag{1}\label{E1}%
\end{equation}
From Table I we can see that the listed four carbides are all stable because
their elastic constants satisfy formula (\ref{E1}). One can find that among
these compounds RuC has the highest bulk modulus with the value $\sim$347 GPa,
which is comparable with that of cubic BN (367 GPa\cite{BN}). The PtC has a
theoretical bulk modulus of 288.7 GPa, in good agreement with the experimental
value of 301 GPa. Thus Table I\ shows that the rocksalt RuC, PdC, together
with PtC, form a new family of low compressible compounds.

We notice that the trends of bulk modulus in these carbides are in some way
related to the number of valence electrons. This can be seen by the fact that
the value of bulk modulus decreases with increasing the total number of
valence electrons (12 per primitive cell for RuC, 14 for PdC and PtC, and 15
for AgC). To further investigate the origin of the bulk modulus, we have
calculated the partial DOS of these carbides. The results are shown in Figs.
2.%
\begin{figure}[tbp]
\begin{center}
\includegraphics[width=1.0\linewidth]{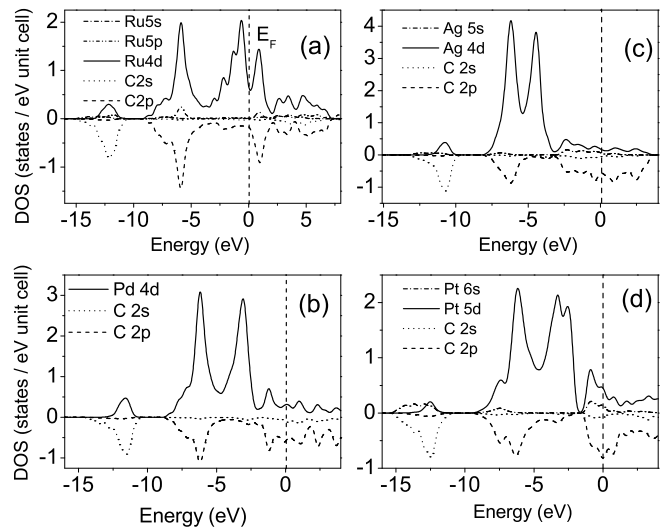}
\end{center}
\caption
{Partial DOS of (a) RuC, (b) PdC, (c) AgC and (d) PtC with rock-salt structure.
}
\label{fig2}
\end{figure}
For RuC (Fig. 2a), there exists a minimum in DOS near the Fermi level, giving
the value of about 0.6 states/eV per primitive cell. The states between --15
eV and --11 eV are mainly composed of C (2\textit{s}) states with a little
contribution from Ru (4\textit{d}) states. There exist two peaks of the Ru
(4\textit{d}) states between --9 eV and the Fermi level. The strong hybridized
states of Ru (4d) and C (2\textit{p}) states between --9 eV and --4 eV is
responsible for the covalent bonding of RuC. For PdC and PtC which contains
two more valence electrons than RuC, the extra electron fills the anti-bonding
\textit{d} orbital of transition metal atoms, thereby decreasing the bonding
strength of such compounds. As a consequence, the bulk modulus of PdC and PtC
are much lower than that of RuC. The same rule can apply to AgC. One may ask
the question why there is an invisible difference of bulk modulus between PdC
and PtC. This can be explained as follows: the presence of 4\textit{f}
electrons in the latter compound, which should repel \textit{d}-orbitals out
of core regions thus enhance the \textit{d}-\textit{p} bonding (see the DOS
patterns of Fig. 2b and Fig. 2c), giving a higher value of bulk modulus for
PtC. In another way for a quantitative analysis, we have investigated the
Fermi energy position of PtC, PdC and AgC compared to RuC, whose Fermi energy
locates at \textit{d} states of transition metal atoms. We find the more the
Fermi level separated from \textit{d} states, the lower the bulk modulus
approaches. The trend is compatible with above proposed mechanism for bulk
modulus decreasement. Furthermore, from Figs. 2 one can conclude that all
these four carbides are metallic.

As mentioned above, the tetragonal shear modulus $G^{\prime\prime}$ measures
the response of a crystal to the volume-conserving tetragonal shear strain. It
is thought that the stretching of metal-C bonds and the bending of metal-metal
bonds are involved in such procedure\cite{wu}. Because $c_{11}$ is determined
by the nearest-neighbor interaction as well as bulk modulus \textit{B}, thus
when \textit{c}$_{11}$%
$>$%
$>$%
\textit{c}$_{12}$, $G^{\prime\prime}$ have the same trend as \textit{B}, as we
observed in RuC and PdC in Table I. However, when $c_{11}$ is comparable to
\textit{c}$_{12}$, the shear modulus trend does not need to be similar as that
of bulk modulus. For example, in the present calculations, the bulk modulus of
PtC is 288.7, much larger than that of AgC (158.8 GPa). However, its
$G^{\prime\prime}$ (27.8 GPa) is smaller than that of AgC (42.7 GPa).

The nature of hardness has been extensively investigated and many new models
have been proposed\cite{tete,Liu,petr,gao,he}. Although for the strong
covalent materials, the hardness can be directly evaluated\cite{gao,he}, for
the partially covalent transition-metal based hard materials, the
polycrystalline shear modulus $G$ [$(c_{11}-c_{12}+3c_{44})/5$] is still
considered as the most important parameter besides the bulk modulus, governing
the indentation hardness. Our calculated polycrystalline shear modulus $G$ of
RuC is 65.2 GPa, which is the highest value of these four noble metal
carbides. This value is just comparable to that of pure Pt (61
GPa\cite{brandes}), remarkably contrasting the very high value of bulk
modulus. It was also found that the highest shear modulus \textit{c}$_{44}%
$\ of these four compounds (PtC: 49.4 GPa) is much lower than that of
conventional materials. For example, the \textit{c}$_{44}$ of pure Pt is 76.4
GPa.\cite{brandes} Therefore these four potential noble metal carbides are not
expected to withstand shear strain to a large extend and may have very low hardness.

In conclusion, we have performed systematic \textit{ab initio} total energy
calculations on the elastic stability of all the rock-salt structured noble
metal carbides. It was found that four carbides, namely RuC, PdC, AgC and PtC,
are mechanically stable. Most of these potential transition metal carbides
have low compressibility though their hardness is not expected to be high. The
low compressibility was found to be related to the occupancy of the valence
states. From the calculated electronic structure, all these carbides is
metallic. Since the rock-slat PtC has been recently synthesized under high
pressure and high temperature, we expect the other three carbides can also be
prepared under certain extreme conditions.

This work was supported by the National Natural Science Foundation of China
(Grant Nos. 10404035,10534030\&50325103).

\end{document}